\documentclass{article}

 \usepackage{neurips_2025}

\usepackage[utf8]{inputenc} 
\usepackage[T1]{fontenc}    
\usepackage{hyperref}       
\usepackage{url}            
\usepackage{booktabs}       
\usepackage{amsfonts}       
\usepackage{nicefrac}       
\usepackage{microtype}      
\usepackage{xcolor}         

\title{Rethinking the Simulation vs. Rendering Dichotomy: No Free Lunch in Spatial World Modelling}

\author{%
  Dezhi Luo \\
  University of Michigan\\
  \texttt{ihzedoul@umich.edu} \\
  \And
  Qingying Gao \\
  Johns Hopkins University\\
  \texttt{qgao14@jh.edu} \\
  \And
  Hokin Deng \\
  Carnegie Mellon University\\
  \texttt{hokind@andrew.cmu.edu} \\
}

\begin{document}

\maketitle

\begin{abstract}

Spatial world models, representations that support flexible reasoning about spatial relations, are central to developing computational models that could operate in the physical world, but their precise mechanistic underpinnings are nuanced by the borrowing of underspecified or misguided accounts of human cognition. This paper revisits the simulation versus rendering dichotomy and draws on evidence from aphantasia to argue that fine-grained perceptual content is critical for model-based spatial reasoning. Drawing on recent research into the neural basis of visual awareness, we propose that spatial simulation and perceptual experience depend on shared representational geometries captured by higher-order indices of perceptual relations. We argue that recent developments in embodied AI support this claim, where rich perceptual details improve performance on physics-based world engagements. To this end, we call for the development of architectures capable of maintaining structured perceptual representations as a step toward spatial world modelling in AI.
  
\end{abstract}

\section{Introduction}

Human reasoning is distinguished by its capacity to construct structured simulations of the external world, or world models, that support prediction, planning, and counterfactual inference beyond what is immediately perceived \citep{johnson1983mental, wooldridge1995intelligent, lecun2022path}. Among these, \textbf{spatial world models} serve as a foundational substrate for structured spatial reasoning: the ability to represent and manipulate spatial relations in a flexible, generalizable manner \citep{kuipers1978modeling, kosslyn1979demystification, hegarty2011cognitive, gao2025vision}. Recently, spatial world modelling has garnered significant attention as a promising approach for enhancing spatial reasoning capabilities in foundation models \citep{chaplot2021differentiable, spies2024transformers, gao2025vision, cai2025has}. Yet, the underlying computational mechanisms remain poorly understood, in part because the concept of a spatial world model is often treated as an idealized abstraction rather than an empirically grounded construct.

Current evaluation methods for model-based spatial reasoning tend to rely on task paradigms that are assumed to require internal simulation, such as mental rotation, visual perspective-taking, and mechanical reasoning \citep{jelassi2022vision, gao2024vision, sun2024probing, zhang2024vision,wang2025vision}. These assumptions are often informed by cognitive science frameworks that describe how humans approach spatial problems \citep{shepard1971mental, hegarty2004mechanical}. If a task is \textit{believed} to necessitate mental simulation of spatial transformations, it is categorized as diagnostic of model-based reasoning. However, such paradigms tend to ignore the nuanced nature of the internal mechanisms humans employ to navigate these problems, which await consensus among cognitive scientists. 


This paper addresses this important challenge of formulating a clear conceptualization of spatial world models: namely, what kinds of internal representations are required to sustain robust simulations of the external world’s spatial properties? To this end, we first revisit the so-called “simulation vs. rendering” dichotomy in mental imagery research. Examining the case of aphantasia, we argue that the standard formulation of this dichotomy is misconstrued. Rather than treating simulation and rendering as separate processes \citep{balaban2025physics}, we propose that both may rely on convergent, higher-order representations that capture the instantiated relations between perceptual objects. The implication is that the representational substrate of spatial world models must be more fine-grained than previously assumed, requiring architectures that go beyond shallow modelling of spatial relations.

\section{On the Split Between Simulation and Rendering}

Efforts to characterize spatial world models often appeal to how humans appear to solve tasks that presumably require internal simulation. A canonical example is mental rotation, where response times scale with angular disparity, suggesting incremental manipulation of internal spatial representations \citep{pylyshyn1979rate, khooshabeh2013individual, hilton2022general}. This has been interpreted as evidence for simulation-based reasoning \citep{pylyshyn1979rate, searle2017mental}. However, the underlying mechanisms are not directly observable, and mental rotation is typically accompanied by \textbf{conscious visual experience}, raising the question of whether such imagery plays a functional role or is merely epiphenomenal \citep{wexler1998motor, vingerhoets2002motor}. To address this, recent proposals distinguish between \textbf{physics-based} and \textbf{graphics-based} components of mental imagery \citep{balaban2025physics}. On this view, tasks such as rotation, scanning, and manipulation may be solved through \textit{simulations over structured spatial representations, without requiring perceptual rendering}. Rendering may still occur, but it bares no direct relevance to correctly and robustly completing these tasks. Here, we note that this position hinges on a linear interpretation of the functional specialization across streams of visual processing in the human brain, and that it is misguided. Below, we discuss the theoretical claims and corresponding evidence in objection to the above interpretation, and highlight that they instead point toward a \textbf{non-linear, hierarchical relationship} between simulation and rendering.

\subsection{Examining the Spatial Imagery Framework and the Case of Aphantasia}

The proposed simulation vs. rendering dichotomy draws particular support from studies of \textbf{aphantasia}. Individuals with aphantasia report an inability to generate voluntary visual images, yet often perform comparably to the general population on tasks such as mental rotation \citep{kay2024slower, balaban2025physics}. This has been interpreted as evidence that spatial simulation can operate independently of conscious vision. To explain this, researchers have proposed a distinction between \textbf{spatial imagery }and visual or object-based imagery, with the former supporting abstract reasoning about relative positions, distances, and spatial relations without relying on pictorial content. On this account, preserved task performance in aphantasia is attributed to intact spatial imagery mechanisms. \citet{balaban2025physics} further has broadened this concept to encompass physical simulation for reasoning about motion and interaction. Under this conceptualization, spatial imagery encodes the entire scene—including geometric structure, spatial extent, and other object properties—sufficient to simulate its temporal evolution. According to them, while no rendering occurs, non-physical attributes can still be abstractly represented, akin to an engineer’s amodal model. Importantly, this implies that simulations required for capturing the spatial relations of the real-world environment, sufficient for structured physical reasoning, do not demand the computations used to support the \textbf{fine-grained perceptual content} in the conscious visual scene that someone without aphantasia would experience when completing tasks.

While the spatial imagery framework offers a plausible account of preserved task performance in aphantasia \citep{kay2024slower}, it rests on a notion that remains under-specified. Notably, it is not apparent what exactly constitutes the content of spatial "imagery".  Proponents of the framework often suggest that spatial imagery involves abstracted or schematic representations of spatial features, such as location, relation, or shape, that are "modality-neutral" and potentially amodal. \citet{phillips2025aphantasia}, for example, describes such imagery as “neutral as to whether the location, relation, shape or structure is imagined as seen or touched.” Nevertheless, when describing how such abstract "imageries" are deployed in action, an example is given as the "subjects imagine grasping the shape and rotating it". This framing seems to rely on the covert recruitment of embodied sensory modalities without clarifying how the spatial content itself is accessed or represented. Simply put, there does not seem to be any reason to insist that the spatial component underlying such recruitment must be consciously experienced. 

If this commitment to conscious experience in the spatial imagery framework is denied, the account then aligns more closely with theories proposing that aphantasia reflects preserved unconscious mental imagery \citep{michel2025aphantasia}. Yet this position, too, has faced criticism with respect to the notion of "imagery". Some argue that imagery inherently entails fine-grained perceptual content, typically associated with modality-specific sensory representations—for instance, those involving early visual cortex \citep{scholz2025absence}. From that perspective, if individuals with aphantasia lack the capacity to reconstruct such content, it is unclear how spatial imagery, defined in amodal terms, could qualify as imagery at all regardless of its experiential status. Given the dubious nature of the spatial imagery framework with respect to phenomonology, we suggest that a more tractable position may be to accept that individuals with aphantasia rely on spatial representations—whether imagistic or not—to solve tasks such as mental rotation, while lacking the corresponding visual experience \citep{hinton1979imagery, nanay2021unconscious}. Determining whether such representations count as “imagery” may ultimately be undecidable, and for the purpose of conceptualizing spatial world models, this ambiguity is largely orthogonal. 

\subsection{Rethinking the Linear Interpretation}

Now that we have clarified the nuances surrounding the aphantasia debate, a key question for understanding the mechanistic basis of spatial world modelling is what kinds of spatial representations underlie task performance in individuals with aphantasia. In this context, \citet{balaban2025physics} propose that aphantasia serves as a proof-of-concept for their claim that spatial/physical imagery and visual/object imagery, acting as alleged instantiations of simulation and rendering, are subserved by distinct mechanisms, a view grounded in the neuroanatomical dissociation between the \textbf{dorsal} and \textbf{ventral} visual streams. The dorsal stream is associated with action-oriented spatial processing, while the ventral stream supports object recognition and visual imagery \citep{goodale1992separate}. However, this \textbf{linear interpretation} of functional specialization is highly contested. Accumulating evidence from inter-stream crosstalk, perceptual integration, and lesion studies suggests that the two streams are not strictly independent, but instead operate within a more distributed and interactive network architecture \citep{schenk2010we, de2011usefulness,milner2017two}. This is not to say that all functional specializations with respect to the dorsal and ventral streams are non-linear, and what  \citet{balaban2025physics} did  acknowledge this nuance and maintain otherwise. However, we highlight that simulation and rendering is indeed one of these examples where this interpretation does not hold, precisely due to the role of spatial representations in conscious vision. 

Following the conceptualization in the original linear interpretation, the dorsal stream has traditionally been cast as the “zombie” stream—so named for its presumed lack of contribution to conscious vision \citep{goodale1992separate, chalmers1997conscious, de2011usefulness, milner2012visual, wu2014against}. However, later accounts challenge this view by proposing that dorsal-stream information related to spatial encoding may in fact play a direct role in shaping visual experience. To begin with, converging evidence from lesion studies and neuroimaging data highlights the involvement of intraparietal regions such as the ventral intraparietal area (VIP) and the lateral intraparietal area (LIP), which encode head- and body-centered reference frames and integrate corollary discharge signals. These dorsal mechanisms help maintain stable spatial representations, including perceived distance, across saccades and object motion, thereby anchoring egocentric spatial frameworks that are essential to the continuity of visual experience \citep{wu2014against}.

This reappraisal has been substantiated by neural evidence indicating that the high-level regions of the dorsal stream is part of the broader fronto-parietal network that encodes fine-grained perceptual content integral to conscious vision. Rather than merely representing coarse spatial representations to support action-oriented or post-perceptual executive functions, these regions appear to participate directly in constructing the contents of visual experience by interacting with the ventral stream, suggesting that simulation and rendering may not be functionally dissociated. In particular, it has been demonstrated that both prefrontal and posterior parietal cortices can decode object identity from rapidly presented visual stimuli even in the absence of behavioral report, with decoders performing above chance in the posterior parietal cortex (PPC), a region traditionally associated with the dorsal “zombie” stream \citep{bellet2022decoding}. This finding indicates that dorsal areas actively encode perceptual information rather than merely mediating visuomotor transformations. Going further upstream, the dorsolateral prefrontal cortex (DLPFC), a region long implicated in leveraging spatial encodings for action planning and cognitive control, has also been shown to selectively correlate with visually aware information \citep{lau2006relative,anzulewicz2019bringing}. Complementary evidence from lesion and studies further supports this view: damage to or deactivation of prefrontal and parietal cortices impairs the integration and maintenance of visual content in awareness \citep{szczepanski2014insights, persaud2011awareness}, while prefrontal ensembles exhibit category- and configuration-selective tuning that evolves with visual experience \citep{rainer2000effects, chan2013functional}. Collectively, these results suggest that simulation and perceptual representation are deeply intertwined within the fronto-parietal network. Rather than functioning as isolated control or sensory systems, the dorsal and ventral streams appear to interact dynamically, integrating spatial encodings with object-level perceptual details to sustain a unified neurocognitive substrate for conscious vision \citep{panagiotaropoulos2024integrative, rees2007neural}.

Together, these findings provide empirical support for higher-order theories (HOT) of consciousness, which posit that meta-representations of perceptual content, supported by fronto-parietal networks, play a central role in shaping perceptual awareness. In particular, relational HOT—including perceptual reality monitoring (PRM) \citep{lau2019consciousness} and higher-order state space (HOSS) \citep{fleming2020awareness} models—converge on the view that the representational substrates of higher-order encodings underlies the phenomenal character of experience, including fine-grained spatial properties such as perceived distances and object relations \citep{fleming2024quality}. On these views, higher-order representations encode how perceptual states relate to one another within structured quality spaces. Crucially, these higher-order representations are not themselves inherently conscious. Rather, they must be discriminated as either reliable reflections of external reality or internally generated signals (e.g., imagination or noise) in order to be gated into conscious awareness. PRM, for instance, likens this gating mechanism to a discriminator in a generative adversarial network (GAN), whereby higher-order processes evaluate which sensory signals are “real” enough to be experienced \citep{gershman2019generative, lau2022consciousness}. Upon making such a determination, the discriminator generates a pointer that carries information about the location where the totality of visual details within experience is stored. This pointer is then fed back into first-order networks for decoding, thereby rendering the conscious experience \citep{lau2019consciousness, butlin2023consciousness}. In this view, while rendering itself (the decoding process) may not directly contribute to spatial reasoning, the fine‑grained geometric structure embedded within higher‑order indices that capture complex object-level properties can suffice the kinds of internal simulations underlying spatial world modelling.

This framework assumes a \textbf{non-linear, hierarchical relationship between simulation and rendering}: the capacity to simulate physical dynamics and the capacity for conscious visual experience may rely on the same underlying representational substrate—differing only in whether these states are verified and utilized by downstream systems \citep{rosenthal2010think, fleming2024quality}. From this perspective, broken rendering in aphantasia does not indicate a failure of a dedicated visual stream, but rather a failure in the gating or feedback process: either the discriminator fails to classify a state as a sufficiently fine-grained and reliable index of reality versus imagination, or the downstream consumer system fails to register and decode the output \citep{michel2025aphantasia}. Both failures impact the conscious awareness of mental simulations but not its functional role in solving spatial reasoning tasks. In the case of aphantasia, where performance on tasks demanding model-based spatial reasoning remains intact, it is more likely to be the latter. This is because top-down determination of the reliability of first-order information remains essential for decision-making, even if such processes do not generate a conscious experience. Supporting this view, a recent study of aphantasia patients found that all 12 cases of lesion-induced aphantasia involved damage to regions functionally connected to the left fusiform gyrus, a region strongly implicated in visual mental imagery. Notably, no significant lesions were found in prefrontal cortices, suggesting that higher-order monitoring mechanisms remained intact \citep{kutsche2025visual}. This pattern implies that the deficit in aphantasia may arise from impaired downstream decoding, rather than from the absence of higher-order representations themselves. Taken together, these findings suggest that fine-grained perceptual representations underlying the content of conscious vision are intrinsic to the functional architecture supporting model-based spatial reasoning in humans, going against the alleged dichotomy between simulation and rendering.

\section{No Free Lunch in Spatial World Modelling}

Rather than treating physical simulation and mental imagery as outputs of dissociable visual systems, the non-linear interpretation locates the critical bottleneck in higher-order re-representation and discrimination processes—mechanisms responsible for indexing and gating perceptual content into conscious experience. This view is consistent with both the broader evidence of dorsal–ventral integration and the specific lesion patterns observed in cases of aphantasia. Crucially, this matters beyond human neuroscience: we propose that the so-called spatial world models in AI may rely on the very same class of structured, fine-grained perceptual representations that underlie conscious visual experience in humans.

In short, there is \textit{no free lunch} in spatial modeling: if human spatial reasoning depends on fine-grained encodings, we should not expect coarse, perceptually abstract approximations to suffice for models aiming to emulate such capacities. This aligns with recent evolutionary accounts proposing that conscious visual experience evolved to stabilize internal simulations—allowing organisms to determine when to commit to a world model and act. On this view, consciousness may have arisen as a by-product mechanism that gates simulations with sufficient fidelity to guide decision-making in the wild \citep{fleming2025sensory}. Importantly, we do not claim that the specific computational implementations underlying human spatial cognition constitute the only viable path toward such capabilities in artificial systems. Nevertheless, they remain the only empirically validated example of an architecture that demonstrably supports such capacities, and hence are worth considering as a principled reference point for developing spatially-grounded AI \citep{cassenti2022editor, zador2022toward,li2024core,luo2025philosophical}. Below, we show that analogous directions have begun to emerge across recent advances in AI and outline future prospects for such paradigms.

\subsection{Language Models Are Not Spatially Competent}

Multimodal language models (MLLMs) \citep{li2023blip2, liu2024visual, team2023gemini} acquire rich visual priors by aligning linguistic and visual data, enabling them to generate coherent spatial descriptions without direct training on visuospatial tasks \citep{ashutosh2025llmsheartraining, sharma2024visioncheckuplanguagemodels, dengreinforcement}. These models demonstrate impressive performance in perception and high-level reasoning \citep{fu2023mme, yang2025magic}, yet they continue to struggle with tasks that require model-based spatial reasoning, including mental rotation, perspective-taking, and mechanical reasoning \citep{gao2024vision, sun2024probing, zhang2024vision, luo2024vision, li2024core, lievaluating, sun2025probing, wang2025vision, cai2025has}, indicating a lack of structured spatial representations and dynamic transformation mechanisms essential for genuine spatial understanding. Although it has been proposed that, as models scale across tasks and modalities, their latent spaces converge toward shared statistical abstractions of the external world \citep{huh2024platonicrepresentationhypothesis}, this convergence cannot occur meaningfully in the spatial domain without vehicles capable of encoding perceptually rich representations. 

\subsection{Implicit Models for Embodied Control}

Building physics engines  as explicit spatial world models has been a holy grail in robotics. MuJoCo enabled precise physics and DRL/LfD breakthroughs in manipulation with encouraging sim-to-real transfer; GPU-based physics engines like Isaac Gym and Genesis scale this via massive parallelism \citep{todorov2012mujoco,kumar2015mujoco,rajeswaran2017learning,gupta2019relay,zhu2019dexterous,zhu2020ingredients,makoviychuk2021isaac,authors2024genesis}. Yet excelling in physics simulation rarely yields structured real-world understanding; policies often lack sufficient models for counterfactuals, and concept grounding, leading to brittleness in open-ended, visually complex settings. In addition, physics-engine–only model-predictive control is brittle: model–reality gaps, partial observability, and heavy contact-dynamics compute force short-horizon, over-tuned policies that transfer poorly \citep{zhang2025whole, pezzato2025sampling}.

Recent vision foundation models (VFMs) has led to remarkable features such as extracting structured features from high-resolution images, improving reasoning and action \citep{simeoni2025dinov3}. Their scaling-driven, modality-agnostic latents align with language and proprioception, enabling multimodal fusion, generative imagination, and possibly spatiotemporal structure \citep{luo2025visionlanguagemodelscreatecrossmodal,huh2024platonicrepresentationhypothesis,chern2025thinking,Raugel2024Disentangling}. These representational benefits translate directly into improved control. Integrating vision foundation models as perceptual modules within robotic learning pipelines has yielded dramatic performance gains—not only in basic manipulation tasks but also in long-horizon planning and generalization \citep{majumdar2023we}. A striking example is the VIP framework, which uses vision-based embeddings to encode task rewards for DRL, effectively reshaping policy optimization through perceptual guidance \citep{ma2022vip}. Such approaches enable more efficient policy learning and produce agents whose behavior transfers exceptionally well from simulation to the real world \citep{shah2021rrl, ma2023liv, chen2023visual, nair2022r3m}. This body of work has led to the emergence of a broader paradigm often referred to as visual pretraining for manipulation, in which vision encoders pretrained on internet-scale image and video data are reused to accelerate downstream control. These pretrained representations serve as implicit world models—embedding structured spatial priors that enhance sample efficiency, generalization, and robustness in policy learning \citep{du2019implicit, du2023learning, zhou2024robodreamer, zhou2025learning}.

Together, in the context of embodied AI, it is evident that both explicit and implicit world models have proven indispensable. Simulation engines such as MuJoCo, Isaac Gym, and Genesis provide detailed physical environments for training agents or model-predictive control \citep{todorov2012mujoco, makoviychuk2021isaac, authors2024genesis}, but these remain insufficient on their own for achieving robust, real-world performance. By contrast, implicit world models, exemplified by frameworks such as VIP and R3M, leverage pretrained visual representations for control generalizations \citep{ma2022vip, nair2022r3m}.

\subsection{Video Models for Action Imagination}

A central question is whether models can, much like humans during navigation, perceive, model, and render spatially relevant tasks without language. On the perceptual side, vision foundation models (e.g., VGG, DINO, CLIP) already exhibit strong transfer across diverse downstream tasks \citep{mei2025geometrymeetsvisionrevisiting, siméoni2025dinov3}. Recent work such as VGGT grounds large vision backbones in a 3D reconstruction objective, yielding visual–geometry priors that capture fine structural detail and boost performance on spatially demanding tasks \citep{wang2025vggtvisualgeometrygrounded}. Inspired by chain-of-thought in language models, we hypothesize that \emph{chain-of-frames} rollouts (i.e., video generation) can enable step-by-step spatiotemporal reasoning. Early evidence comes from Veo~3, which generates next-scene predictions conditioned on video inputs to tackle visual reasoning tasks such as Sudoku, mazes, and navigation \citep{wiedemer2025videomodelszeroshotlearners}. We can leverage pretrained video diffusion models to synthesize video-based plans for actions \citep{du2023learning, yang2024video}. We see an exciting opportunity in new approaches such as Genex and Genie-3 \citep{lu2024genex, lu2024generative, deepmind_genie3_2025} and see huge potential in them to mature into brain-like implicit world models, surpassing 2D pixel representations by internalizing scene geometry, dynamics, and affordances. Humans “run movies in the mind” before acting, using vivid imagery to simulate candidate futures. Analogously, we see that video models can generate action rollouts that score and refine action sequences, supplying policies with strong visuomotor plans. We hypothesize that this imagination-as-video strategy could let machines reason about space as we do.

\section{Conclusion}

This paper revisited the long-standing debate between simulation and rendering by drawing on insights from both cognitive neuroscience and embodied AI. We argued that human spatial reasoning is unlikely to rely on purely schematic or amodal simulations; instead, it depends on perceptually rich content that stabilizes internal models of the world. Recent developments in embodied AI and robotics reinforce this view: models that leverage large-scale vision encoders to provide high-fidelity embeddings, serving as implicit world models, could outperform those trained solely through physics-based simulation. These findings mirror how humans integrate detailed perceptual information into mental simulations, drawing on shared representational structures rather than separate systems. Closing the gap between human and artificial spatial reasoning may therefore depend on leveraging rich perceptual grounding in structured spatial representations.

\bibliographystyle{plainnat}
\bibliography{references}

\begin{thebibliography}{103}
\providecommand{\natexlab}[1]{#1}
\providecommand{\url}[1]{\texttt{#1}}
\expandafter\ifx\csname urlstyle\endcsname\relax
  \providecommand{\doi}[1]{doi: #1}\else
  \providecommand{\doi}{doi: \begingroup \urlstyle{rm}\Url}\fi

\bibitem[Anzulewicz et~al.(2019)Anzulewicz, Hobot, Siedlecka, and Wierzcho{\'n}]{anzulewicz2019bringing}
Anna Anzulewicz, Justyna Hobot, Marta Siedlecka, and Micha{\l} Wierzcho{\'n}.
\newblock Bringing action into the picture. how action influences visual awareness.
\newblock \emph{Attention, Perception, \& Psychophysics}, 81\penalty0 (7):\penalty0 2171--2176, 2019.

\bibitem[Ashutosh et~al.(2025)Ashutosh, Gandelsman, Chen, Misra, and Girdhar]{ashutosh2025llmsheartraining}
Kumar Ashutosh, Yossi Gandelsman, Xinlei Chen, Ishan Misra, and Rohit Girdhar.
\newblock Llms can see and hear without any training, 2025.
\newblock URL \url{https://arxiv.org/abs/2501.18096}.

\bibitem[Balaban and Ullman(2025)]{balaban2025physics}
Halely Balaban and Tomer~D Ullman.
\newblock Physics versus graphics as an organizing dichotomy in cognition.
\newblock \emph{Trends in Cognitive Sciences}, 2025.

\bibitem[Bellet et~al.(2022)Bellet, Gay, Dwarakanath, Jarraya, Van~Kerkoerle, Dehaene, and Panagiotaropoulos]{bellet2022decoding}
Joachim Bellet, Marion Gay, Abhilash Dwarakanath, Bechir Jarraya, Timo Van~Kerkoerle, Stanislas Dehaene, and Theofanis~I Panagiotaropoulos.
\newblock Decoding rapidly presented visual stimuli from prefrontal ensembles without report nor post-perceptual processing.
\newblock \emph{Neuroscience of consciousness}, 2022\penalty0 (1):\penalty0 niac005, 2022.

\bibitem[Butlin et~al.(2023)Butlin, Long, Elmoznino, Bengio, Birch, Constant, Deane, Fleming, Frith, Ji, et~al.]{butlin2023consciousness}
Patrick Butlin, Robert Long, Eric Elmoznino, Yoshua Bengio, Jonathan Birch, Axel Constant, George Deane, Stephen~M Fleming, Chris Frith, Xu~Ji, et~al.
\newblock Consciousness in artificial intelligence: insights from the science of consciousness.
\newblock \emph{arXiv preprint arXiv:2308.08708}, 2023.

\bibitem[Cai et~al.(2025)Cai, Wang, Sun, Wang, Gu, Yin, Lin, Yang, Wei, Shi, et~al.]{cai2025has}
Zhongang Cai, Yubo Wang, Qingping Sun, Ruisi Wang, Chenyang Gu, Wanqi Yin, Zhiqian Lin, Zhitao Yang, Chen Wei, Xuanke Shi, et~al.
\newblock Has gpt-5 achieved spatial intelligence? an empirical study.
\newblock \emph{arXiv preprint arXiv:2508.13142}, 2025.

\bibitem[Cassenti et~al.(2022)Cassenti, Veksler, and Ritter]{cassenti2022editor}
Daniel~N Cassenti, Vladislav~D Veksler, and Frank~E Ritter.
\newblock Editor's review and introduction: Cognition-inspired artificial intelligence, 2022.

\bibitem[Chalmers(1997)]{chalmers1997conscious}
David~J Chalmers.
\newblock \emph{The conscious mind: In search of a fundamental theory}.
\newblock Oxford Paperbacks, 1997.

\bibitem[Chan(2013)]{chan2013functional}
Annie W-Y Chan.
\newblock Functional organization and visual representations of human ventral lateral prefrontal cortex.
\newblock \emph{Frontiers in psychology}, 4:\penalty0 371, 2013.

\bibitem[Chaplot et~al.(2021)Chaplot, Pathak, and Malik]{chaplot2021differentiable}
Devendra~Singh Chaplot, Deepak Pathak, and Jitendra Malik.
\newblock Differentiable spatial planning using transformers.
\newblock In \emph{International conference on machine learning}, pages 1484--1495. PMLR, 2021.

\bibitem[Chen et~al.(2023)Chen, Tippur, Wu, Kumar, Adelson, and Agrawal]{chen2023visual}
Tao Chen, Megha Tippur, Siyang Wu, Vikash Kumar, Edward Adelson, and Pulkit Agrawal.
\newblock Visual dexterity: In-hand reorientation of novel and complex object shapes.
\newblock \emph{Science Robotics}, 8\penalty0 (84):\penalty0 eadc9244, 2023.

\bibitem[Chern et~al.(2025)Chern, Hu, Chern, Kou, Su, Ma, Deng, and Liu]{chern2025thinking}
Ethan Chern, Zhulin Hu, Steffi Chern, Siqi Kou, Jiadi Su, Yan Ma, Zhijie Deng, and Pengfei Liu.
\newblock Thinking with generated images.
\newblock \emph{arXiv preprint arXiv:2505.22525}, 2025.

\bibitem[de~Haan and Cowey(2011)]{de2011usefulness}
Edward~HF de~Haan and Alan Cowey.
\newblock On the usefulness of ‘what’and ‘where’pathways in vision.
\newblock \emph{Trends in cognitive sciences}, 15\penalty0 (10):\penalty0 460--466, 2011.

\bibitem[Deng(2025)]{dengreinforcement}
Hokin Deng.
\newblock Reinforcement learning versus natural language programs: Where is flexible planning and problem solving in natural intelligence coming from?
\newblock \emph{OSF}, 2025.

\bibitem[Du and Mordatch(2019)]{du2019implicit}
Yilun Du and Igor Mordatch.
\newblock Implicit generation and modeling with energy based models.
\newblock \emph{Advances in neural information processing systems}, 32, 2019.

\bibitem[Du et~al.(2023)Du, Yang, Dai, Dai, Nachum, Tenenbaum, Schuurmans, and Abbeel]{du2023learning}
Yilun Du, Sherry Yang, Bo~Dai, Hanjun Dai, Ofir Nachum, Josh Tenenbaum, Dale Schuurmans, and Pieter Abbeel.
\newblock Learning universal policies via text-guided video generation.
\newblock \emph{Advances in neural information processing systems}, 36:\penalty0 9156--9172, 2023.

\bibitem[Fleming(2020)]{fleming2020awareness}
Stephen~M Fleming.
\newblock Awareness as inference in a higher-order state space.
\newblock \emph{Neuroscience of consciousness}, 2020\penalty0 (1):\penalty0 niz020, 2020.

\bibitem[Fleming and Michel(2025)]{fleming2025sensory}
Stephen~M Fleming and Matthias Michel.
\newblock Sensory horizons and the functions of conscious vision.
\newblock \emph{Behavioral and Brain Sciences}, pages 1--53, 2025.

\bibitem[Fleming and Shea(2024)]{fleming2024quality}
Stephen~M Fleming and Nicholas Shea.
\newblock Quality space computations for consciousness.
\newblock \emph{Trends in Cognitive Sciences}, 28\penalty0 (10):\penalty0 896--906, 2024.

\bibitem[Fu et~al.(2023)Fu, Chen, Shen, Qin, Zhang, Lin, Yang, Zheng, Li, Sun, Wu, and Ji]{fu2023mme}
Chaoyou Fu, Peixian Chen, Yunhang Shen, Yulei Qin, Mengdan Zhang, Xu~Lin, Jinrui Yang, Xiawu Zheng, Ke~Li, Xing Sun, Yunsheng Wu, and Rongrong Ji.
\newblock Mme: A comprehensive evaluation benchmark for multimodal large language models.
\newblock \emph{arXiv preprint arXiv: 2306.13394}, 2023.

\bibitem[Gao et~al.(2024)Gao, Li, Lyu, Sun, Luo, and Deng]{gao2024vision}
Qingying Gao, Yijiang Li, Haiyun Lyu, Haoran Sun, Dezhi Luo, and Hokin Deng.
\newblock Vision language models see what you want but not what you see.
\newblock \emph{arXiv preprint arXiv:2410.00324}, 2024.

\bibitem[Gao et~al.(2025)Gao, Pi, Liu, Chen, Yang, Huang, Fang, Sun, Kishore, Ai, et~al.]{gao2025vision}
Qiyue Gao, Xinyu Pi, Kevin Liu, Junrong Chen, Ruolan Yang, Xinqi Huang, Xinyu Fang, Lu~Sun, Gautham Kishore, Bo~Ai, et~al.
\newblock Do vision-language models have internal world models? towards an atomic evaluation.
\newblock \emph{arXiv preprint arXiv:2506.21876}, 2025.

\bibitem[Gemini(2023)]{team2023gemini}
Gemini.
\newblock Gemini: A family of highly capable multimodal models.
\newblock \emph{arXiv preprint arXiv: 2312.11805}, 2023.

\bibitem[Gershman(2019)]{gershman2019generative}
Samuel~J Gershman.
\newblock The generative adversarial brain.
\newblock \emph{Frontiers in Artificial Intelligence}, 2:\penalty0 18, 2019.

\bibitem[Goodale and Milner(1992)]{goodale1992separate}
Melvyn~A Goodale and A~David Milner.
\newblock Separate visual pathways for perception and action.
\newblock \emph{Trends in neurosciences}, 15\penalty0 (1):\penalty0 20--25, 1992.

\bibitem[Gupta et~al.(2019)Gupta, Kumar, Lynch, Levine, and Hausman]{gupta2019relay}
Abhishek Gupta, Vikash Kumar, Corey Lynch, Sergey Levine, and Karol Hausman.
\newblock Relay policy learning: Solving long-horizon tasks via imitation and reinforcement learning.
\newblock \emph{arXiv preprint arXiv:1910.11956}, 2019.

\bibitem[Hegarty(2004)]{hegarty2004mechanical}
Mary Hegarty.
\newblock Mechanical reasoning by mental simulation.
\newblock \emph{Trends in cognitive sciences}, 8\penalty0 (6):\penalty0 280--285, 2004.

\bibitem[Hegarty(2011)]{hegarty2011cognitive}
Mary Hegarty.
\newblock The cognitive science of visual-spatial displays: Implications for design.
\newblock \emph{Topics in cognitive science}, 3\penalty0 (3):\penalty0 446--474, 2011.

\bibitem[Hilton et~al.(2022)Hilton, Raddatz, and Gramann]{hilton2022general}
Christopher Hilton, Leonie Raddatz, and Klaus Gramann.
\newblock A general spatial transformation process? assessing the neurophysiological evidence on the similarity of mental rotation and folding.
\newblock \emph{Neuroimage: Reports}, 2\penalty0 (2):\penalty0 100092, 2022.

\bibitem[Hinton(1979)]{hinton1979imagery}
Geoffrey Hinton.
\newblock Imagery without arrays.
\newblock \emph{Behavioral and Brain Sciences}, 2\penalty0 (4):\penalty0 555--556, 1979.

\bibitem[Huh et~al.(2024)Huh, Cheung, Wang, and Isola]{huh2024platonicrepresentationhypothesis}
Minyoung Huh, Brian Cheung, Tongzhou Wang, and Phillip Isola.
\newblock The platonic representation hypothesis, 2024.
\newblock URL \url{https://arxiv.org/abs/2405.07987}.

\bibitem[Jelassi et~al.(2022)Jelassi, Sander, and Li]{jelassi2022vision}
Samy Jelassi, Michael Sander, and Yuanzhi Li.
\newblock Vision transformers provably learn spatial structure.
\newblock \emph{Advances in Neural Information Processing Systems}, 35:\penalty0 37822--37836, 2022.

\bibitem[Johnson-Laird(1983)]{johnson1983mental}
Philip~Nicholas Johnson-Laird.
\newblock \emph{Mental models: Towards a cognitive science of language, inference, and consciousness}.
\newblock Number~6. Harvard University Press, 1983.

\bibitem[Kay et~al.(2024)Kay, Keogh, and Pearson]{kay2024slower}
Lachlan Kay, Rebecca Keogh, and Joel Pearson.
\newblock Slower but more accurate mental rotation performance in aphantasia linked to differences in cognitive strategies.
\newblock \emph{Consciousness and cognition}, 121:\penalty0 103694, 2024.

\bibitem[Khooshabeh et~al.(2013)Khooshabeh, Hegarty, and Shipley]{khooshabeh2013individual}
Peter Khooshabeh, Mary Hegarty, and Thomas~F Shipley.
\newblock Individual differences in mental rotation.
\newblock \emph{Experimental psychology}, 2013.

\bibitem[Kosslyn et~al.(1979)Kosslyn, Pinker, Smith, and Shwartz]{kosslyn1979demystification}
Stephen~M Kosslyn, Steven Pinker, George~E Smith, and Steven~P Shwartz.
\newblock On the demystification of mental imagery.
\newblock \emph{Behavioral and Brain Sciences}, 2\penalty0 (4):\penalty0 535--548, 1979.

\bibitem[Kuipers(1978)]{kuipers1978modeling}
Benjamin Kuipers.
\newblock Modeling spatial knowledge.
\newblock \emph{Cognitive science}, 2\penalty0 (2):\penalty0 129--153, 1978.

\bibitem[Kumar and Todorov(2015)]{kumar2015mujoco}
Vikash Kumar and Emanuel Todorov.
\newblock Mujoco haptix: A virtual reality system for hand manipulation.
\newblock In \emph{2015 IEEE-RAS 15th International Conference on Humanoid Robots (Humanoids)}, pages 657--663. IEEE, 2015.

\bibitem[Kutsche et~al.(2025)Kutsche, Howard, Drew, Michel, Cohen, Fox, and Kletenik]{kutsche2025visual}
Julian Kutsche, Calvin Howard, William Drew, Matthias Michel, Alexander~L Cohen, Michael~D Fox, and Isaiah Kletenik.
\newblock Visual mental imagery and aphantasia lesions map onto a convergent brain network.
\newblock \emph{medRxiv}, pages 2025--05, 2025.

\bibitem[Lau(2019)]{lau2019consciousness}
Hakwan Lau.
\newblock Consciousness, metacognition, \& perceptual reality monitoring.
\newblock \emph{PsyArXiv}, 2019.

\bibitem[Lau(2022)]{lau2022consciousness}
Hakwan Lau.
\newblock \emph{In consciousness we trust: The cognitive neuroscience of subjective experience}.
\newblock Oxford University Press, 2022.

\bibitem[Lau and Passingham(2006)]{lau2006relative}
Hakwan~C Lau and Richard~E Passingham.
\newblock Relative blindsight in normal observers and the neural correlate of visual consciousness.
\newblock \emph{Proceedings of the National Academy of Sciences}, 103\penalty0 (49):\penalty0 18763--18768, 2006.

\bibitem[LeCun(2022)]{lecun2022path}
Yann LeCun.
\newblock A path towards autonomous machine intelligence version 0.9. 2, 2022-06-27.
\newblock \emph{Open Review}, 62\penalty0 (1):\penalty0 1--62, 2022.

\bibitem[Li et~al.(2023)Li, Li, Savarese, and Hoi]{li2023blip2}
Junnan Li, Dongxu Li, Silvio Savarese, and Steven Hoi.
\newblock Blip-2: Bootstrapping language-image pre-training with frozen image encoders and large language models.
\newblock \emph{CONFERENCE}, 2023.

\bibitem[Li et~al.(2024)Li, Gao, Zhao, Wang, Sun, Lyu, Hawkins, Vasconcelos, Golan, Luo, et~al.]{li2024core}
Yijiang Li, Qingying Gao, Tianwei Zhao, Bingyang Wang, Haoran Sun, Haiyun Lyu, Robert~D Hawkins, Nuno Vasconcelos, Tal Golan, Dezhi Luo, et~al.
\newblock Core knowledge deficits in multi-modal language models.
\newblock \emph{arXiv preprint arXiv:2410.10855}, 2024.

\bibitem[Li et~al.(2025)Li, Wang, Zhao, Gao, Deng, and Luo]{lievaluating}
Yijiang Li, Bingyang Wang, Tianwei Zhao, Qingying Gao, Hokin Deng, and Dezhi Luo.
\newblock Evaluating multi-modal language models through concept hacking.
\newblock In \emph{Workshop on Spurious Correlation and Shortcut Learning: Foundations and Solutions}, 2025.

\bibitem[Liu et~al.(2024)Liu, Li, Wu, and Lee]{liu2024visual}
Haotian Liu, Chunyuan Li, Qingyang Wu, and Yong~Jae Lee.
\newblock Visual instruction tuning.
\newblock \emph{Advances in neural information processing systems}, 36, 2024.

\bibitem[Lu et~al.(2024{\natexlab{a}})Lu, Shu, Xiao, Ye, Wang, Peng, Wei, Khashabi, Chellappa, Yuille, et~al.]{lu2024genex}
Taiming Lu, Tianmin Shu, Junfei Xiao, Luoxin Ye, Jiahao Wang, Cheng Peng, Chen Wei, Daniel Khashabi, Rama Chellappa, Alan Yuille, et~al.
\newblock Genex: Generating an explorable world.
\newblock \emph{arXiv preprint arXiv:2412.09624}, 2024{\natexlab{a}}.

\bibitem[Lu et~al.(2024{\natexlab{b}})Lu, Shu, Yuille, Khashabi, and Chen]{lu2024generative}
Taiming Lu, Tianmin Shu, Alan Yuille, Daniel Khashabi, and Jieneng Chen.
\newblock Generative world explorer.
\newblock \emph{arXiv preprint arXiv:2411.11844}, 2024{\natexlab{b}}.

\bibitem[Luo et~al.(2024)Luo, Lyu, Gao, Sun, Li, and Deng]{luo2024vision}
Dezhi Luo, Haiyun Lyu, Qingying Gao, Haoran Sun, Yijiang Li, and Hokin Deng.
\newblock Vision language models know law of conservation without understanding more-or-less.
\newblock \emph{arXiv preprint arXiv:2410.00332}, 2024.

\bibitem[Luo et~al.(2025{\natexlab{a}})Luo, Li, and Deng]{luo2025philosophical}
Dezhi Luo, Yijiang Li, and Hokin Deng.
\newblock The philosophical foundations of growing ai like a child.
\newblock \emph{arXiv preprint arXiv:2502.10742}, 2025{\natexlab{a}}.

\bibitem[Luo et~al.(2025{\natexlab{b}})Luo, Darrell, and Bar]{luo2025visionlanguagemodelscreatecrossmodal}
Grace Luo, Trevor Darrell, and Amir Bar.
\newblock Vision-language models create cross-modal task representations, 2025{\natexlab{b}}.
\newblock URL \url{https://arxiv.org/abs/2410.22330}.

\bibitem[Ma et~al.(2022)Ma, Sodhani, Jayaraman, Bastani, Kumar, and Zhang]{ma2022vip}
Yecheng~Jason Ma, Shagun Sodhani, Dinesh Jayaraman, Osbert Bastani, Vikash Kumar, and Amy Zhang.
\newblock Vip: Towards universal visual reward and representation via value-implicit pre-training.
\newblock \emph{arXiv preprint arXiv:2210.00030}, 2022.

\bibitem[Ma et~al.(2023)Ma, Kumar, Zhang, Bastani, and Jayaraman]{ma2023liv}
Yecheng~Jason Ma, Vikash Kumar, Amy Zhang, Osbert Bastani, and Dinesh Jayaraman.
\newblock Liv: Language-image representations and rewards for robotic control.
\newblock In \emph{International Conference on Machine Learning}, pages 23301--23320. PMLR, 2023.

\bibitem[Majumdar et~al.(2023)Majumdar, Yadav, Arnaud, Ma, Chen, Silwal, Jain, Berges, Wu, Vakil, et~al.]{majumdar2023we}
Arjun Majumdar, Karmesh Yadav, Sergio Arnaud, Jason Ma, Claire Chen, Sneha Silwal, Aryan Jain, Vincent-Pierre Berges, Tingfan Wu, Jay Vakil, et~al.
\newblock Where are we in the search for an artificial visual cortex for embodied intelligence?
\newblock \emph{Advances in Neural Information Processing Systems}, 36:\penalty0 655--677, 2023.

\bibitem[Makoviychuk et~al.(2021)Makoviychuk, Wawrzyniak, Guo, Lu, Storey, Macklin, Hoeller, Rudin, Allshire, Handa, et~al.]{makoviychuk2021isaac}
Viktor Makoviychuk, Lukasz Wawrzyniak, Yunrong Guo, Michelle Lu, Kier Storey, Miles Macklin, David Hoeller, Nikita Rudin, Arthur Allshire, Ankur Handa, et~al.
\newblock Isaac gym: High performance gpu-based physics simulation for robot learning.
\newblock \emph{arXiv preprint arXiv:2108.10470}, 2021.

\bibitem[Mei et~al.(2025)Mei, Shorinwa, and Majumdar]{mei2025geometrymeetsvisionrevisiting}
Zhiting Mei, Ola Shorinwa, and Anirudha Majumdar.
\newblock Geometry meets vision: Revisiting pretrained semantics in distilled fields, 2025.
\newblock URL \url{https://arxiv.org/abs/2510.03104}.

\bibitem[Michel et~al.(2025)Michel, Morales, Block, and Lau]{michel2025aphantasia}
Matthias Michel, Jorge Morales, Ned Block, and Hakwan Lau.
\newblock Aphantasia as imagery blindsight.
\newblock \emph{Trends in Cognitive Sciences}, 2025.

\bibitem[Milner(2012)]{milner2012visual}
A~David Milner.
\newblock Is visual processing in the dorsal stream accessible to consciousness?
\newblock \emph{Proceedings of the Royal Society B: Biological Sciences}, 279\penalty0 (1737):\penalty0 2289--2298, 2012.

\bibitem[Milner(2017)]{milner2017two}
A~David Milner.
\newblock How do the two visual streams interact with each other?
\newblock \emph{Experimental brain research}, 235\penalty0 (5):\penalty0 1297--1308, 2017.

\bibitem[Nair et~al.(2022)Nair, Rajeswaran, Kumar, Finn, and Gupta]{nair2022r3m}
Suraj Nair, Aravind Rajeswaran, Vikash Kumar, Chelsea Finn, and Abhinav Gupta.
\newblock R3m: A universal visual representation for robot manipulation.
\newblock \emph{arXiv preprint arXiv:2203.12601}, 2022.

\bibitem[Nanay(2021)]{nanay2021unconscious}
Bence Nanay.
\newblock Unconscious mental imagery.
\newblock \emph{Philosophical Transactions of the Royal Society B}, 376\penalty0 (1817):\penalty0 20190689, 2021.

\bibitem[Panagiotaropoulos(2024)]{panagiotaropoulos2024integrative}
Theofanis~I Panagiotaropoulos.
\newblock An integrative view of the role of prefrontal cortex in consciousness.
\newblock \emph{Neuron}, 112\penalty0 (10):\penalty0 1626--1641, 2024.

\bibitem[Parker-Holder and Fruchter()]{deepmind_genie3_2025}
Jack Parker-Holder and Shlomi Fruchter.
\newblock Genie 3: A new frontier for world models.
\newblock URL \url{https://deepmind.google/discover/blog/genie-3-a-new-frontier-for-world-models/}.
\newblock Blog post.

\bibitem[Persaud et~al.(2011)Persaud, Davidson, Maniscalco, Mobbs, Passingham, Cowey, and Lau]{persaud2011awareness}
Navindra Persaud, Matthew Davidson, Brian Maniscalco, Dean Mobbs, Richard~E Passingham, Alan Cowey, and Hakwan Lau.
\newblock Awareness-related activity in prefrontal and parietal cortices in blindsight reflects more than superior visual performance.
\newblock \emph{Neuroimage}, 58\penalty0 (2):\penalty0 605--611, 2011.

\bibitem[Pezzato et~al.(2025)Pezzato, Salmi, Trevisan, Spahn, Alonso-Mora, and Corbato]{pezzato2025sampling}
Corrado Pezzato, Chadi Salmi, Elia Trevisan, Max Spahn, Javier Alonso-Mora, and Carlos~Hern{\'a}ndez Corbato.
\newblock Sampling-based model predictive control leveraging parallelizable physics simulations.
\newblock \emph{IEEE Robotics and Automation Letters}, 2025.

\bibitem[Phillips(2025)]{phillips2025aphantasia}
Ian Phillips.
\newblock Aphantasia reimagined.
\newblock \emph{No{\^u}s}, 2025.

\bibitem[Pylyshyn(1979)]{pylyshyn1979rate}
Zenon~W Pylyshyn.
\newblock The rate of “mental rotation” of images: A test of a holistic analogue hypothesis.
\newblock \emph{Memory \& cognition}, 7\penalty0 (1):\penalty0 19--28, 1979.

\bibitem[Rainer and Miller(2000)]{rainer2000effects}
Gregor Rainer and Earl~K Miller.
\newblock Effects of visual experience on the representation of objects in the prefrontal cortex.
\newblock \emph{Neuron}, 27\penalty0 (1):\penalty0 179--189, 2000.

\bibitem[Rajeswaran et~al.(2017)Rajeswaran, Kumar, Gupta, Vezzani, Schulman, Todorov, and Levine]{rajeswaran2017learning}
Aravind Rajeswaran, Vikash Kumar, Abhishek Gupta, Giulia Vezzani, John Schulman, Emanuel Todorov, and Sergey Levine.
\newblock Learning complex dexterous manipulation with deep reinforcement learning and demonstrations.
\newblock \emph{arXiv preprint arXiv:1709.10087}, 2017.

\bibitem[Raugel et~al.(2025)Raugel, Szafraniec, Vo, Couprie, Labatut, Bojanowski, Wyart, and King]{Raugel2024Disentangling}
Joséphine Raugel, Marc Szafraniec, Huy~V. Vo, Camille Couprie, Patrick Labatut, Piotr Bojanowski, Valentin Wyart, and Jean-Rémi King.
\newblock Disentangling the factors of convergence between brains and computer vision models.
\newblock \emph{Manuscript}, page 1–15, 2025.
\newblock preprint PDF, 15 pages. URL: \url{https://scontent-lga3-2.xx.fbcdn.net/v/t39.2365-6/533250329_1837870417159359_3262810533067032733_n.pdf?…}.

\bibitem[Rees(2007)]{rees2007neural}
Geraint Rees.
\newblock Neural correlates of the contents of visual awareness in humans.
\newblock \emph{Philosophical Transactions of the Royal Society B: Biological Sciences}, 362\penalty0 (1481):\penalty0 877--886, 2007.

\bibitem[Rosenthal(2010)]{rosenthal2010think}
David Rosenthal.
\newblock How to think about mental qualities.
\newblock \emph{Philosophical Issues}, 20:\penalty0 368--393, 2010.

\bibitem[Schenk and McIntosh(2010)]{schenk2010we}
Thomas Schenk and Robert~D McIntosh.
\newblock Do we have independent visual streams for perception and action?
\newblock \emph{Cognitive Neuroscience}, 1\penalty0 (1):\penalty0 52--62, 2010.

\bibitem[Scholz et~al.(2025)Scholz, Monzel, and Liu]{scholz2025absence}
Christian~O Scholz, Merlin Monzel, and Jianghao Liu.
\newblock Absence of shared representation in the visual cortex challenges unconscious imagery in aphantasia.
\newblock \emph{Current Biology}, 35\penalty0 (13):\penalty0 R645--R646, 2025.

\bibitem[Searle and Hamm(2017)]{searle2017mental}
Jordan~A Searle and Jeff~P Hamm.
\newblock Mental rotation: An examination of assumptions.
\newblock \emph{Wiley Interdisciplinary Reviews: Cognitive Science}, 8\penalty0 (6):\penalty0 e1443, 2017.

\bibitem[Shah and Kumar(2021)]{shah2021rrl}
Rutav Shah and Vikash Kumar.
\newblock Rrl: Resnet as representation for reinforcement learning.
\newblock \emph{arXiv preprint arXiv:2107.03380}, 2021.

\bibitem[Sharma et~al.(2024)Sharma, Shaham, Baradad, Fu, Rodriguez-Munoz, Duggal, Isola, and Torralba]{sharma2024visioncheckuplanguagemodels}
Pratyusha Sharma, Tamar~Rott Shaham, Manel Baradad, Stephanie Fu, Adrian Rodriguez-Munoz, Shivam Duggal, Phillip Isola, and Antonio Torralba.
\newblock A vision check-up for language models, 2024.
\newblock URL \url{https://arxiv.org/abs/2401.01862}.

\bibitem[Shepard and Metzler(1971)]{shepard1971mental}
Roger~N Shepard and Jacqueline Metzler.
\newblock Mental rotation of three-dimensional objects.
\newblock \emph{Science}, 171\penalty0 (3972):\penalty0 701--703, 1971.

\bibitem[Sim{\'e}oni et~al.(2025)Sim{\'e}oni, Vo, Seitzer, Baldassarre, Oquab, Jose, Khalidov, Szafraniec, Yi, Ramamonjisoa, et~al.]{simeoni2025dinov3}
Oriane Sim{\'e}oni, Huy~V Vo, Maximilian Seitzer, Federico Baldassarre, Maxime Oquab, Cijo Jose, Vasil Khalidov, Marc Szafraniec, Seungeun Yi, Micha{\"e}l Ramamonjisoa, et~al.
\newblock Dinov3.
\newblock \emph{arXiv preprint arXiv:2508.10104}, 2025.

\bibitem[Siméoni et~al.(2025)Siméoni, Vo, Seitzer, Baldassarre, Oquab, Jose, Khalidov, Szafraniec, Yi, Ramamonjisoa, Massa, Haziza, Wehrstedt, Wang, Darcet, Moutakanni, Sentana, Roberts, Vedaldi, Tolan, Brandt, Couprie, Mairal, Jégou, Labatut, and Bojanowski]{siméoni2025dinov3}
Oriane Siméoni, Huy~V. Vo, Maximilian Seitzer, Federico Baldassarre, Maxime Oquab, Cijo Jose, Vasil Khalidov, Marc Szafraniec, Seungeun Yi, Michaël Ramamonjisoa, Francisco Massa, Daniel Haziza, Luca Wehrstedt, Jianyuan Wang, Timothée Darcet, Théo Moutakanni, Leonel Sentana, Claire Roberts, Andrea Vedaldi, Jamie Tolan, John Brandt, Camille Couprie, Julien Mairal, Hervé Jégou, Patrick Labatut, and Piotr Bojanowski.
\newblock Dinov3, 2025.
\newblock URL \url{https://arxiv.org/abs/2508.10104}.

\bibitem[Spies et~al.(2024)Spies, Edwards, Ivanitskiy, Skapars, R{\"a}uker, Inoue, Russo, and Shanahan]{spies2024transformers}
Alex~F Spies, William Edwards, Michael~I Ivanitskiy, Adrians Skapars, Tilman R{\"a}uker, Katsumi Inoue, Alessandra Russo, and Murray Shanahan.
\newblock Transformers use causal world models in maze-solving tasks.
\newblock \emph{arXiv preprint arXiv:2412.11867}, 2024.

\bibitem[Sun et~al.(2024)Sun, Gao, Lyu, Luo, Li, and Deng]{sun2024probing}
Haoran Sun, Qingying Gao, Haiyun Lyu, Dezhi Luo, Yijiang Li, and Hokin Deng.
\newblock Probing mechanical reasoning in large vision language models.
\newblock \emph{arXiv preprint arXiv:2410.00318}, 2024.

\bibitem[Sun et~al.(2025)Sun, Yu, Li, Gao, Lyu, Deng, and Luo]{sun2025probing}
Haoran Sun, Suyang Yu, Yijiang Li, Qingying Gao, Haiyun Lyu, Hokin Deng, and Dezhi Luo.
\newblock Probing perceptual constancy in large vision language models.
\newblock \emph{arXiv preprint arXiv:2502.10273}, 2025.

\bibitem[Szczepanski and Knight(2014)]{szczepanski2014insights}
Sara~M Szczepanski and Robert~T Knight.
\newblock Insights into human behavior from lesions to the prefrontal cortex.
\newblock \emph{Neuron}, 83\penalty0 (5):\penalty0 1002--1018, 2014.

\bibitem[Todorov et~al.(2012)Todorov, Erez, and Tassa]{todorov2012mujoco}
Emanuel Todorov, Tom Erez, and Yuval Tassa.
\newblock Mujoco: A physics engine for model-based control.
\newblock In \emph{2012 IEEE/RSJ international conference on intelligent robots and systems}, pages 5026--5033. IEEE, 2012.

\bibitem[Vingerhoets et~al.(2002)Vingerhoets, de~Lange, Vandemaele, Deblaere, and Achten]{vingerhoets2002motor}
Guy Vingerhoets, Floris~P de~Lange, Pieter Vandemaele, Karel Deblaere, and Erik Achten.
\newblock Motor imagery in mental rotation: an fmri study.
\newblock \emph{Neuroimage}, 17\penalty0 (3):\penalty0 1623--1633, 2002.

\bibitem[Wang et~al.(2025{\natexlab{a}})Wang, Li, Zhou, Leong, Zhao, Ye, Deng, Luo, and Vasconcelos]{wang2025vision}
Bingyang Wang, Yijiang Li, Qingyang Zhou, Hui~Yi Leong, Tianwei Zhao, Letian Ye, Hokin Deng, Dezhi Luo, and Nuno Vasconcelos.
\newblock Do vision language models infer human intention without visual perspective-taking? towards a scalable" one-image-probe-all" dataset.
\newblock In \emph{ICML 2025 Workshop on Assessing World Models}, 2025{\natexlab{a}}.

\bibitem[Wang et~al.(2025{\natexlab{b}})Wang, Chen, Karaev, Vedaldi, Rupprecht, and Novotny]{wang2025vggtvisualgeometrygrounded}
Jianyuan Wang, Minghao Chen, Nikita Karaev, Andrea Vedaldi, Christian Rupprecht, and David Novotny.
\newblock Vggt: Visual geometry grounded transformer, 2025{\natexlab{b}}.
\newblock URL \url{https://arxiv.org/abs/2503.11651}.

\bibitem[Wexler et~al.(1998)Wexler, Kosslyn, and Berthoz]{wexler1998motor}
Mark Wexler, Stephen~M Kosslyn, and Alain Berthoz.
\newblock Motor processes in mental rotation.
\newblock \emph{Cognition}, 68\penalty0 (1):\penalty0 77--94, 1998.

\bibitem[Wiedemer et~al.(2025)Wiedemer, Li, Vicol, Gu, Matarese, Swersky, Kim, Jaini, and Geirhos]{wiedemer2025videomodelszeroshotlearners}
Thaddäus Wiedemer, Yuxuan Li, Paul Vicol, Shixiang~Shane Gu, Nick Matarese, Kevin Swersky, Been Kim, Priyank Jaini, and Robert Geirhos.
\newblock Video models are zero-shot learners and reasoners, 2025.
\newblock URL \url{https://arxiv.org/abs/2509.20328}.

\bibitem[Wooldridge and Jennings(1995)]{wooldridge1995intelligent}
Michael Wooldridge and Nicholas~R Jennings.
\newblock Intelligent agents: Theory and practice.
\newblock \emph{The knowledge engineering review}, 10\penalty0 (2):\penalty0 115--152, 1995.

\bibitem[Wu(2014)]{wu2014against}
Wayne Wu.
\newblock Against division: Consciousness, information and the visual streams.
\newblock \emph{Mind \& Language}, 29\penalty0 (4):\penalty0 383--406, 2014.

\bibitem[Yang et~al.(2024)Yang, Walker, Parker-Holder, Du, Bruce, Barreto, Abbeel, and Schuurmans]{yang2024video}
Sherry Yang, Jacob Walker, Jack Parker-Holder, Yilun Du, Jake Bruce, Andre Barreto, Pieter Abbeel, and Dale Schuurmans.
\newblock Video as the new language for real-world decision making.
\newblock \emph{arXiv preprint arXiv:2402.17139}, 2024.

\bibitem[Yang et~al.(2025)Yang, Luo, Han, and Hovy]{yang2025magic}
Shuo Yang, Siwen Luo, Soyeon~Caren Han, and Eduard Hovy.
\newblock Magic-vqa: Multimodal and grounded inference with commonsense knowledge for visual question answering.
\newblock \emph{arXiv preprint arXiv:2503.18491}, 2025.

\bibitem[Zador et~al.(2022)Zador, Escola, Richards, {\"O}lveczky, Bengio, Boahen, Botvinick, Chklovskii, Churchland, Clopath, et~al.]{zador2022toward}
Anthony Zador, Sean Escola, Blake Richards, Bence {\"O}lveczky, Yoshua Bengio, Kwabena Boahen, Matthew Botvinick, Dmitri Chklovskii, Anne Churchland, Claudia Clopath, et~al.
\newblock Toward next-generation artificial intelligence: Catalyzing the neuroai revolution.
\newblock \emph{arXiv preprint arXiv:2210.08340}, 2022.

\bibitem[Zhang et~al.(2025)Zhang, Howell, Yi, Pan, Shi, Qu, Erez, Tassa, and Manchester]{zhang2025whole}
John~Z Zhang, Taylor~A Howell, Zeji Yi, Chaoyi Pan, Guanya Shi, Guannan Qu, Tom Erez, Yuval Tassa, and Zachary Manchester.
\newblock Whole-body model-predictive control of legged robots with mujoco.
\newblock \emph{arXiv preprint arXiv:2503.04613}, 2025.

\bibitem[Zhang et~al.(2024)Zhang, Hu, Lee, Shi, Kordjamshidi, Chai, and Ma]{zhang2024vision}
Zheyuan Zhang, Fengyuan Hu, Jayjun Lee, Freda Shi, Parisa Kordjamshidi, Joyce Chai, and Ziqiao Ma.
\newblock Do vision-language models represent space and how? evaluating spatial frame of reference under ambiguities.
\newblock \emph{arXiv preprint arXiv:2410.17385}, 2024.

\bibitem[Zhou et~al.(2024{\natexlab{a}})Zhou, Du, Chen, Li, Yeung, and Gan]{zhou2024robodreamer}
Siyuan Zhou, Yilun Du, Jiaben Chen, Yandong Li, Dit-Yan Yeung, and Chuang Gan.
\newblock Robodreamer: Learning compositional world models for robot imagination.
\newblock \emph{arXiv preprint arXiv:2404.12377}, 2024{\natexlab{a}}.

\bibitem[Zhou et~al.(2025)Zhou, Du, Yang, Han, Chen, Yeung, and Gan]{zhou2025learning}
Siyuan Zhou, Yilun Du, Yuncong Yang, Lei Han, Peihao Chen, Dit-Yan Yeung, and Chuang Gan.
\newblock Learning 3d persistent embodied world models.
\newblock \emph{arXiv preprint arXiv:2505.05495}, 2025.

\bibitem[Zhou et~al.(2024{\natexlab{b}})Zhou, Qiao, Xu, Wang, Chen, Zheng, Xiong, Wang, Zhang, Ma, Wang, and Dou]{authors2024genesis}
Xian Zhou, Yiling Qiao, Zhenjia Xu, Tsun-Hsuan Wang, Zhehuan Chen, Juntian Zheng, Ziyan Xiong, Yian Wang, Mingrui Zhang, Pingchuan Ma, Yufei Wang, and Zhiyang Dou.
\newblock Genesis: A universal and generative physics engine for robotics and beyond.
\newblock \emph{URL https://github.com/Genesis-Embodied-AI/Genesis}, 2024{\natexlab{b}}.

\bibitem[Zhu et~al.(2019)Zhu, Gupta, Rajeswaran, Levine, and Kumar]{zhu2019dexterous}
Henry Zhu, Abhishek Gupta, Aravind Rajeswaran, Sergey Levine, and Vikash Kumar.
\newblock Dexterous manipulation with deep reinforcement learning: Efficient, general, and low-cost.
\newblock In \emph{2019 International Conference on Robotics and Automation (ICRA)}, pages 3651--3657. IEEE, 2019.

\bibitem[Zhu et~al.(2020)Zhu, Yu, Gupta, Shah, Hartikainen, Singh, Kumar, and Levine]{zhu2020ingredients}
Henry Zhu, Justin Yu, Abhishek Gupta, Dhruv Shah, Kristian Hartikainen, Avi Singh, Vikash Kumar, and Sergey Levine.
\newblock The ingredients of real-world robotic reinforcement learning.
\newblock \emph{arXiv preprint arXiv:2004.12570}, 2020.

\end{thebibliography}





\end{document}